\documentclass{ws-rv9x6}
\usepackage{ws-rv-van}            \makeindex
\begin{document}

\chapter{Magnetic Avalanches in Molecular Nanomagnets}

\author{M. P. Sarachik}

\address{Department of Physics, City College of New York, CUNY,\\ New York, New York, 10031, U. S. A.\\
sarachik@sci.ccny.cuny.edu}

\author[M. P. Sarachik and S. McHugh]{S. McHugh}
\address{Department of Physics, University of California, \\ Santa Barbara, California, 93106-9530, U. S. A.}

\begin{abstract}
The magnetization of the  prototypical molecular magnet Mn$_{12}$-acetate exhibits a series of sharp steps at low temperatures due to quantum tunneling at specific resonant values of magnetic field applied along the easy $c$-axis. An abrupt reversal of the magnetic moment of such a crystal can also occur as an avalanche, where the spin reversal  proceeds along a ``deflagration" front that travels through the sample at subsonic speed.  In this article we review experimental results that have been obtained for the ignition temperature and the speed of propagation of magnetic avalanches in molecular nanomagnets.  Fits of the data with the theory of magnetic deflagration yield overall qualitative agreement.  However, numerical discrepancies indicate that our understanding of these avalanches is incomplete.
\end{abstract}

\body

\section{Background}\label{sec1}
\begin{figure}[h]
\begin{center}
 \parbox{1.5in}{\epsfig{figure=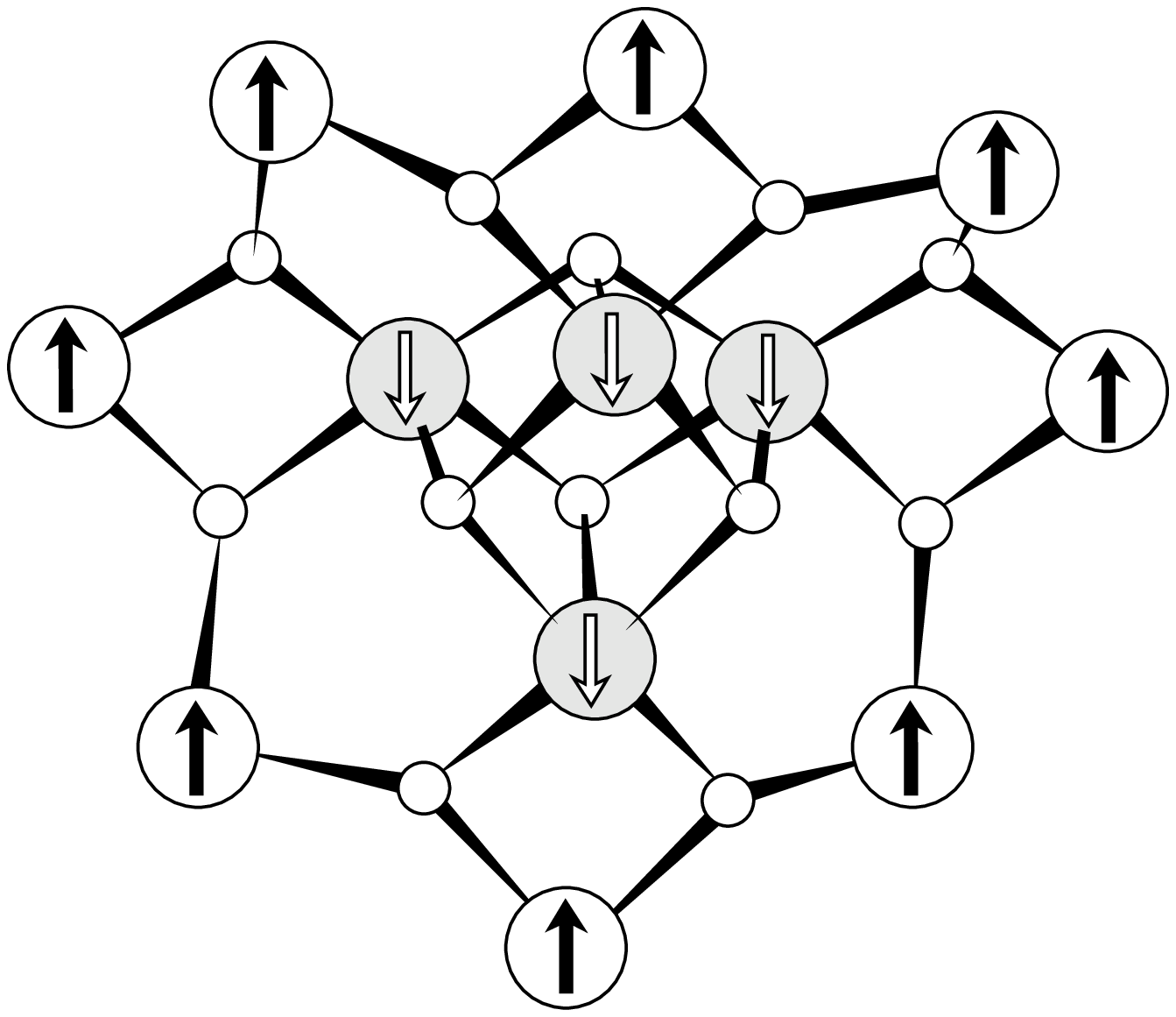,width=1.3in}
  \figsubcap{a}}
  \hspace*{3.0pt}
  \parbox{2.2in}{\epsfig{figure=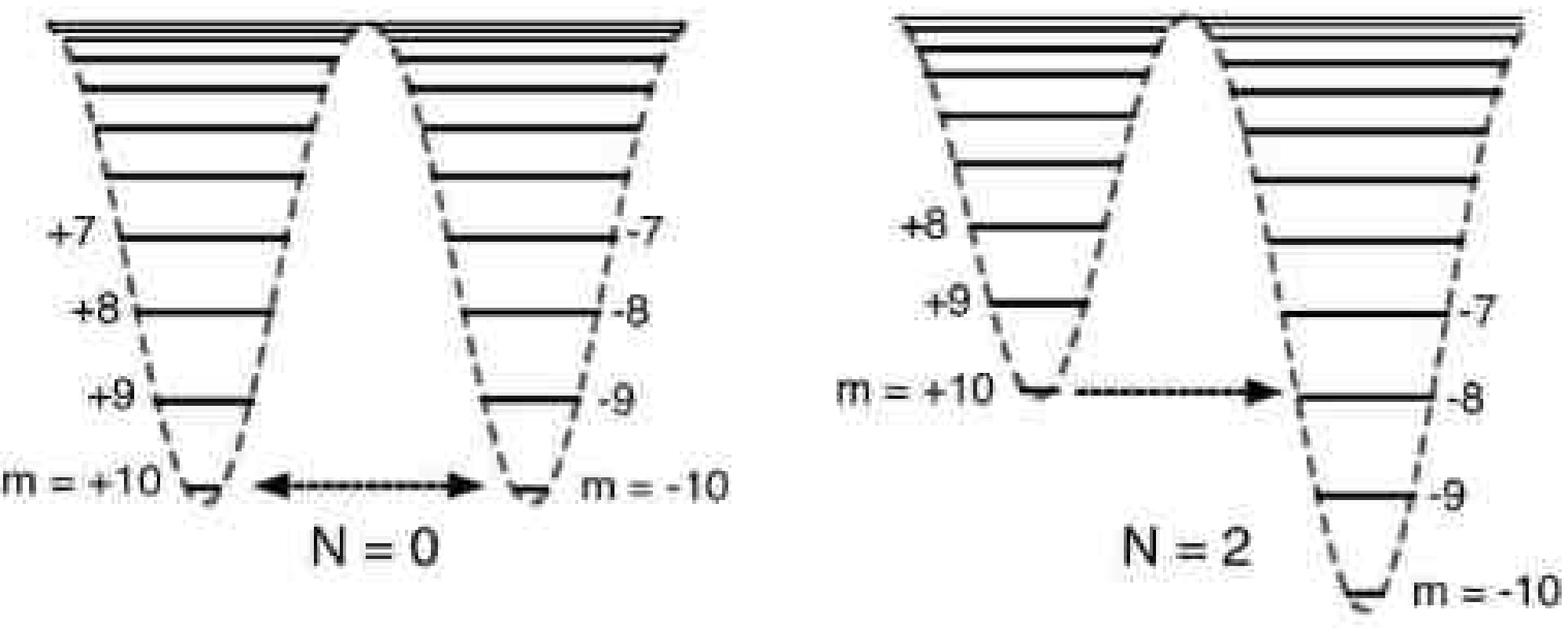,width=2.2in}
  \figsubcap{b}}
  \caption{(a) Chemical structure of the core of the Mn$_{12}$ molecule. The four inner spin-down Mn$^{3+}$ ions each have spin $S=3/2$; the eight outer spin-up Mn$^{4+}$ ions each have spin $S = 2$, yielding a net spin $S = 10$ for the magnetic cluster; the small open circles are O bridges and arrows denote spin. Acetate ligands and water molecules have been removed for clarity. 
  (b) Left: Double-well potential in the absence of magnetic field showing spin-up and spin-down levels separated by the anisotropy barrier. Different spin projection states $|m>$ are indicated. The arrows denote quantum tunneling. Right: Double-well potential for the N=2 step in a magnetic field applied along the easy axis.}
\label{fig1}
\end{center}
\end{figure}

First synthesized by Lis in 1980 \cite{lis}, Mn$_{12}$O$_{12}$(CH$_3$COO)$_{16}$(H$_2$O)$_4$ (referred to hereafter as Mn$_{12}$-ac) is a particularly simple, prototypical molecular magnet.  Shown in Fig.~\ref{fig1}(a), the magnetic core of Mn$_{12}$-ac has four Mn$^{4+}$ (S = 3/2) ions in a central tetrahedron surrounded by eight Mn$^{3+}$ (S = 2) ions. The ions are coupled by superexchange through oxygen bridges with the net result that the four inner and eight outer ions point in opposite directions, yielding a total spin $S=10$ \cite{Sessoli}. The magnetic core is surrounded by acetate ligands, which serve to isolate each core from its neighbors in a body-centered tetragonal lattice.  A crystalline sample contains $\sim 10^{17}$ or more (nearly) identical, weakly interacting single molecule nanomagnets in (nearly) identical crystalline environments.

While there are very weak exchange interactions between molecules, the exchange between ions within the magnetic core is very strong, resulting in a rigid spin $10$ unit that has no internal spin degrees of freedom at low temperatures. To lowest order, the spin Hamiltonian is given by:
\begin{equation}
{\cal H} = - DS_z^2 - g_z\mu_B H_z S_z + \ldots + {\cal H^\prime} 
\label{Hamiltonian}.
\end{equation}
The first term denotes the anisotropy barrier, the second is the Zeeman energy that splits the spin-up and spin-down states in a magnetic field, and the last term, $\cal H^\prime$, contains all symmetry-breaking operators that do not commute with $S_z$.  For Mn$_{12}$-ac, $D=0.548$K, $g_z = 1.94$, and $\mu_B$ is the Bohr magneton.

As illustrated by Figure~\ref{fig1}(b), the spin's energy can be modeled as a double-well potential, where one well corresponds to the spin pointing ``up'' and the other to the spin pointing ``down''. A strong uniaxial anisotropy barrier of the order of 66 K yields doubly degenerate ground states in zero field. The spin has a set of energy levels corresponding to different projections, $m = 10, 9,\ldots, -9, -10$, of the total spin along the easy ($c$-axis) of the crystal \cite{friedmanbook,christoureview,sessolireview,sessolibook}. 

\section{Spin Reversal by Quantum Tunneling}\label{sec2}
\begin{figure}[tb]
\centering
\includegraphics[width=1\linewidth]{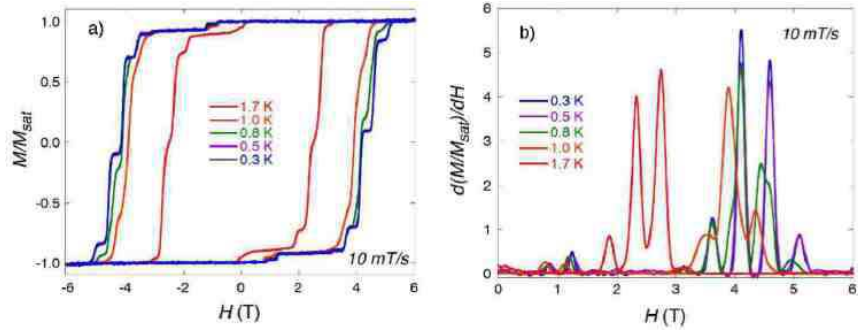}
\caption{(a) Magnetization of a Mn$_{12}$-ac crystal normalized by its saturation value as a function of magnetic field applied along the uniaxial $c$-axis direction at different temperatures below the blocking temperature; the magnetic field was swept at 10 mT/s. (b) The derivative, $dM/dH$ of the data in part (a) as a function of magnetic field.}
\label{loops}
\end{figure}

Slow relaxation below the blocking temperature, $T_B \sim 3$ K, gives rise to hysteresis loops that display steps \cite{Friedman} as a function of magnetic field, $H_z$, swept along the easy $c$-axis of the crystal \cite{TejadaEPL,JMHernandez}.  Figure~\ref{loops}(a) shows the magnetization $M$ as a function of magnetic field $\mu_0H_z$; the derivative, $dM/dH$, which reflects the magnetic relaxation rate, is plotted as a function of $\mu_0H_z$ in Fig.~\ref{loops}(b).  These steps, characteristic of molecular magnets, can be understood with reference to the double well potential of Fig.~\ref{fig1}(b): a magnetic field $H_z$ introduces a Zeeman splitting that tilts the potential wells and causes energy levels in the right (left) well to move down (up). Levels in opposite wells align at particular values of magnetic field (dashed lines in Fig.~\ref{fig1}(b)), allowing the spin to reverse by tunneling.  Full (saturation) magnetization is thereby reached in a stepwise fashion, with the detailed form of the steps depending on sweep-rate and temperature.

%%%%%%%%%%%%%%%%%%%%%%%%%%%%%%%%%%%%%%%%%%%%%%%%%%%
\section{Spin Reversal by Avalanches}\label{sec3}
As first reported by Paulsen and Park \cite{Paulsen},  Mn$_{12}$-ac crystals sometimes exhibit a sudden, complete reversal of magnetic moment during a field-swept measurement.  This phenomenon, also observed in other molecular magnets, was attributed to a thermal runaway (avalanche) in which the relaxation of magnetization toward the direction of the field results in the release of heat that further accelerates the magnetic relaxation.  Direct measurements of the heat emitted have confirmed the thermal nature of the avalanches.  In addition to releasing thermal energy, molecular crystals emit bursts of radiation during magnetic avalanches \cite{tejadaradiation1,tejadaradiation2,keren}.  Once considered events to be avoided, as they interfere with a detailed study of the stepwise process of magnetization, magnetic avalanches became the focus of attention and renewed interest stimulated by the theoretical suggestion that the radiation emitted during an avalanche is in the form of coherent (Dicke) superradiance \cite{superradiance}. Although the issue of coherence of the radiation has yet to be resolved, recent studies have clarified the nature of the avalanche process itself.

\subsection{Magnetic Deflagration}\label{sec3.1}

\begin{figure}[h]
\begin{center}
\includegraphics[width=4 in]{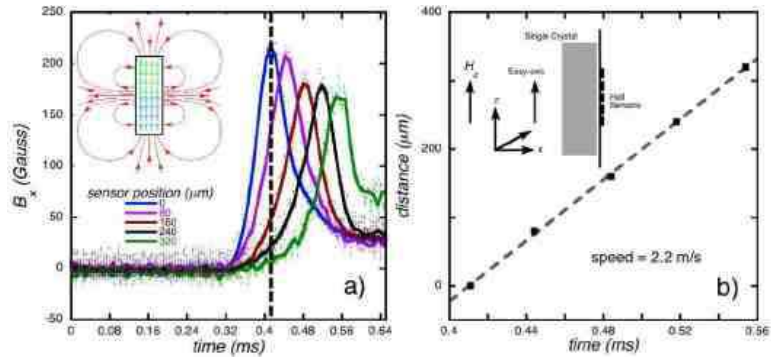}
\caption{(a) The local magnetization measured as a function of time by an array of micron-sized Hall sensors placed along the surface of the sample.  The inset illustrates the ``bunching'' of magnetic field lines as the deflagration front travels past a given Hall sensor.  (b)The sensor position as a function of the time at which the sensor registered the peak; the inset shows the placement of the Hall sensors on the crystal.  The propagation speed for this avalanche is $2.2$ m/s, approximately three orders of magnitude below the speed of sound.}
\end{center}
\label{deflagration}
\end{figure}

From time-resolved measurements of the local magnetization using an array of micron-sized Hall sensors placed on the surface of Mn$_{12}$-ac crystals, Suzuki {\em et al.} \cite{suzuki} discovered that a magnetic avalanche propagates through the crystal at subsonic speed in the form of a thin interface between regions of opposite magnetization.  Figure \ref{deflagration}(a) shows traces recorded during an avalanche by sensors placed in sequential positions near the center of a Mn$_{12}$-ac sample. The inset is a schematic that illustrates the bunching of field lines at the propagating front that gives rise to the observed peaks.  Figure \ref{deflagration}(b) is a plot of the sensor position versus the time of arrival of the peak.  From these measurements one deduces that the front separating up- and down-spins travels with a constant (field-dependent) speed on the order of $5$ m/s, two to three orders of magnitude slower than the speed of sound.

From a thermodynamic point of view, a crystal of Mn$_{12}$ molecules placed in a magnetic field opposite to the magnetic moment is equivalent to a metastable (flammable) chemical  substance.  A well-known mechanism for the release of energy by a metastable chemical substance is combustion or slow burning, technically referred to as deflagration.\cite{LL}  It occurs as a flame front of finite width propagates at a constant speed small compared to the speed of sound.  For ``magnetic deflagration'' in Mn$_{12}$-ac, the role of the chemical energy stored in a molecule is played by the difference in the Zeeman energy, $\Delta E = 2g\mu_BHS$, for states of the Mn$_{12}$-ac molecule that correspond to ${\bf S}$ parallel and antiparallel to ${\bf H}$.

\subsection{Avalanche Ignition}
Although the probability of a spontaneous avalanche has been shown to be higher at resonant magnetic fields than off-resonance \cite{macia}, avalanche ignition is unpredictable and uncontrolled when an external magnetic field is swept back and forth, the experimental protocol generally used to study the steps in the hysteresis loops.  Avalanche ignition under these conditions is a stochastic process that depends on factors such as the sweep rate, the temperature, and the quality of the crystal.  Controlled ignition of avalanches has now been achieved using surface acoustic waves (which serve to heat the sample)  \cite{quantumdeflagration}, and by using a heater \cite{seandips} , as described below and in the next section.

\begin{figure}[h]
\begin{center}
  \parbox{2.0in}{\epsfig{figure=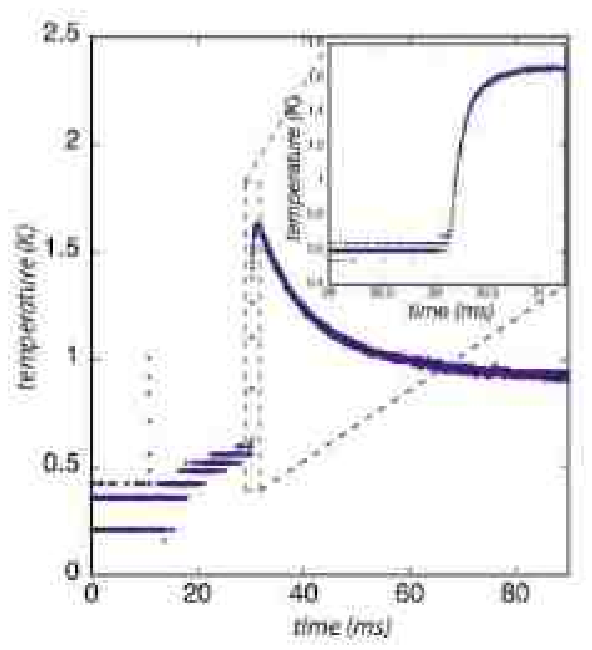,width=1.8in}
  \figsubcap{a}}
  \hspace*{4pt}
  \parbox{2.3in}{\epsfig{figure=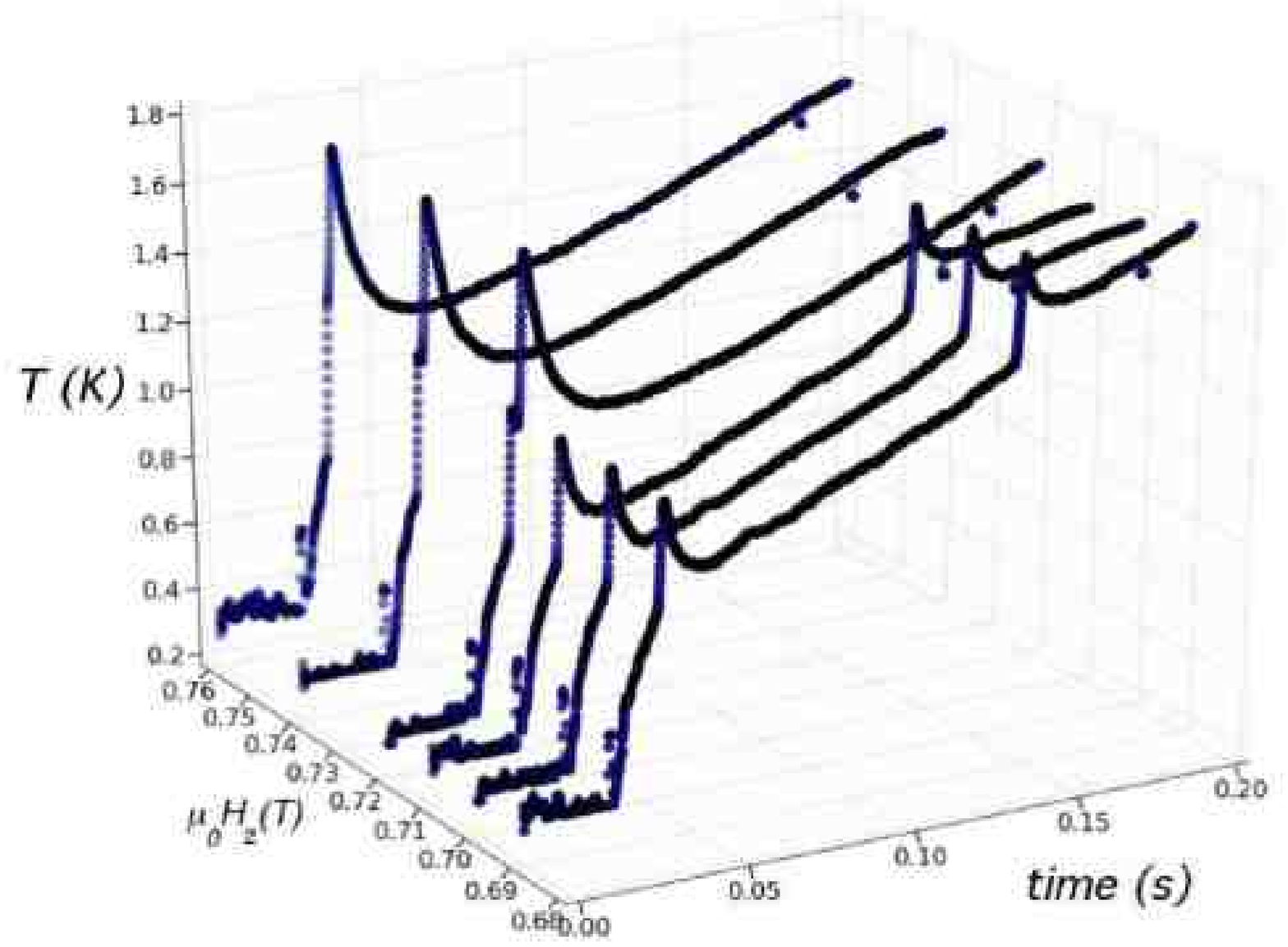,width=2.5in}
  \figsubcap{b}}
  \caption{(a) Temperature recorded by a thermometer in contact with a Mn$_{12}$ crystal during the triggering of an avalanche. The heater is turned on at $\sim 0.012$ s, the temperature then increases slowly until an abrupt rise in temperature at $0.03$ s signals the ignition of an avalanche.  The inset shows data taken near ignition with higher resolution.  The noise at low temperatures derives from digitizing  the analog output of the thermometer, which depends weakly on temperature below $0.4$ K.
 (b) Temperature profiles for avalanches of major and minor species triggered at low fields in a Mn$_{12}$ crystal. The two types of avalanches are triggered separately below a sample-dependent magnetic field, while at higher fields ignition of the minor species triggers the ignition of the major species.}
  \label{ignition}
\end{center}
\end{figure}

McHugh {\em et al.} \cite{seandips} employed a resistive wire element as a simple electric heater to trigger avalanches in a manner similar to the work of Paulsen and Park \cite{Paulsen}.  In these experiments, an external magnetic field is ramped to and held at a fixed value.  The heater is then turned on to slowly heat the sample until an avalanche is triggered at a temperature measured by a small thermometer.  Avalanches launched by this method occur at well-defined, reproducible ignition temperatures.  Fig. \ref{ignition}(a) shows a typical temperature profile: starting at the base temperature of 300 mK, the temperature gradually rises  until an abrupt sharp increase in the temperature signals the ignition of an avalanche.  For this avalanche triggered at $\mu_0 H_z = 1.85$ T, the ignition temperature is about 0.6 K.  

Single crystals of Mn$_{12}$--ac are known to contain two types of molecules. In addition to the primary or ``major'' species described earlier,  as-grown crystals contain a second ``minor'' species at a level of $\approx 5$ percent with lower (rhombohedral) symmetry \cite{minorRef, WernsdorferI}.  These faster-relaxing molecules can be modeled by the same effective spin Hamiltonian, Eq. \ref{Hamiltonian}, with a lower anisotropy barrier of $0.49$ K.   Avalanches of the each species can be studied in the absence of the other through an appropriate magnetic protocol described in Ref. ~\refcite{minors}.   Interestingly, avalanches are separately triggered by the two species in low magnetic field.  As shown in Fig. \ref{ignition}(b), at low fields the minor species relaxes prior to and independently of the major species, while above $\sim 0.7$ T, the major and minor species ignite together and propagate as a single front.  It is analogous to grass and trees that can sustain separate burn fronts that abruptly merge into a single front when the grass becomes sufficiently hot to ignite the trees.

\begin{figure}[tb]
\centering
\includegraphics[width=2.5 in]{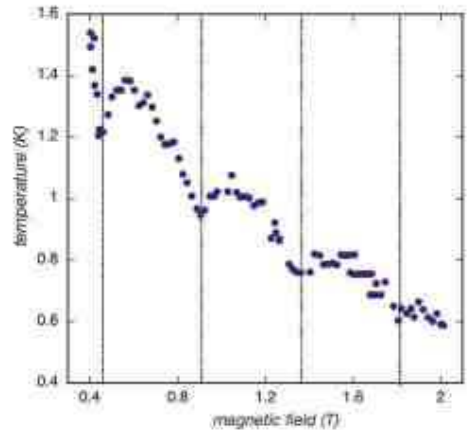}
\caption{Temperature required to ignite avalanches plotted as a function of magnetic field. The vertical lines denote the magnetic fields where sharp minima occur in the ignition temperature corresponding to tunneling near the top of the anisotropy barrier. The overall decrease in ignition temperature is due to the reduction of the anisotropy barrier as the field is increased.}
\label{ignitionminima}
\end{figure}

Despite the turbulent conditions that one might expect for deflagration (analogous to chemical combustion), quantum mechanical tunneling clearly plays a role, as demonstrated in Fig. \ref{ignitionminima}, where the temperature above which ignition occurs is plotted as a function of a preset, constant magnetic field \cite{seandips}.  The temperature required to ignite avalanches exhibits an overall decrease with applied magnetic field, reflecting the fact that larger fields reduce the barrier (see the double-well potential in Fig. \ref{fig1}(b)). The role of quantum mechanics is clearly evidenced by the minima observed for the ignition temperature at the resonant magnetic fields due to tunneling when levels cross on opposite sides of the anisotropy barrier.

\begin{figure}[h]
\begin{center}
  \parbox{2.5in}{\epsfig{figure=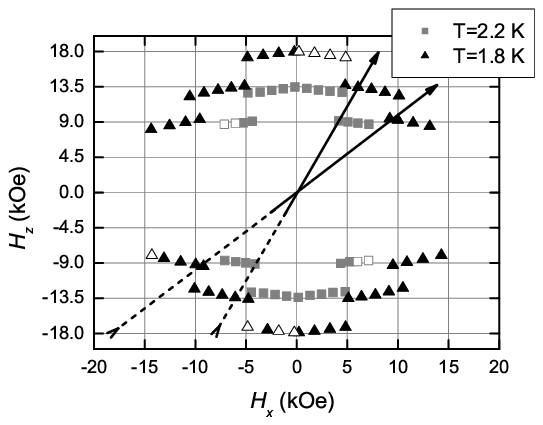,width=2.5in}
  \figsubcap{a}}
  \hspace*{4pt}
  \parbox{1.8in}{\epsfig{figure=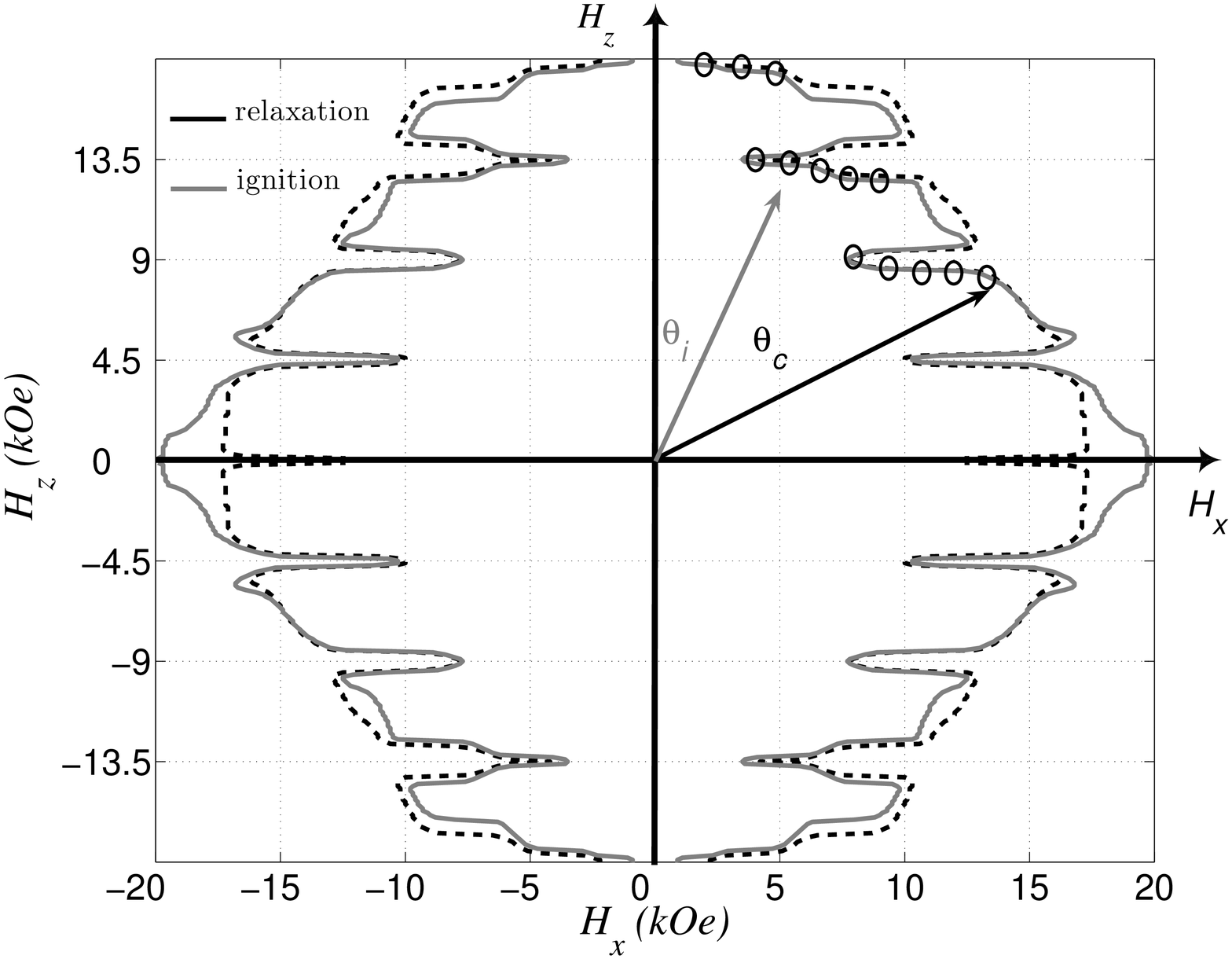,width=2.3in}
  \figsubcap{b}}
\caption{(a) Angle dependence of metastability measured through the occurrence of avalanches.  Squares (triangles) denote parameter values where deflagration occurs for initial temperature 2.2 K (1.8 K).  (b)  Theoretical calculation for the area of stability against ignition of avalanches (solid curve) and against slow relaxation (dashed curve).  Circles denote points where avalanches are predicted to occur at a given angle $\theta_i$ within the first quadrant. The angle $\theta_c$ denotes the crossing point between areas of slow relaxation and avalanche stability.  These results were obtained with T$_f$ as a parameter varying from $6.8$ K for $H = 4600$ Oe to $10.9$ K for $H = 9200$ Oe.  From Maci\`a {\em et al.} \cite{macia}.  }
\label{macia}
\end{center}
\end{figure}

In the ignition studies described above, the barrier against spin reversal was lowered by applying an external magnetic field, $H_z$, along the uniaxial $c$-direction, which serves to unbalance the potential wells and lower the barrier to spin reversal.  A transverse field, $H_x$, also reduces the anisotropy barrier by introducing a symmetry-breaking term, $(g\mu_B H_x S_x) $, to the Hamiltonian, Eq. \ref{Hamiltonian}, thereby promoting tunneling.  Maci\`a {\em et al.} \cite{macia} investigated the threshold for avalanche ignition in Mn$_{12}$-ac as a function of the magnitude and direction of a magnetic field applied at various angles with respect to the anisotropy axis and as a function of temperature.  As the external field is increased at a constant rate from negative saturation to positive values, both $H_z$ and $H_x$ increase, tracing a trajectory in the $(H_z,H_x)$ parameter space; (examples of sweeps starting from zero are shown by the arrows in Fig. \ref{macia}.  As shown in Fig. \ref{macia} (a), an avalanche was recorded for each pair $(H_x, H_z)$ denoted by a square (for $T=2.2$ K) or a triangle (for $T=1.8$) K. We postpone a detailed explanation of these results and a comparison with theory to a later section.

\subsection{Avalanche Speed}

\begin{figure}[tb]
\centering
\includegraphics[width=0.7\linewidth]{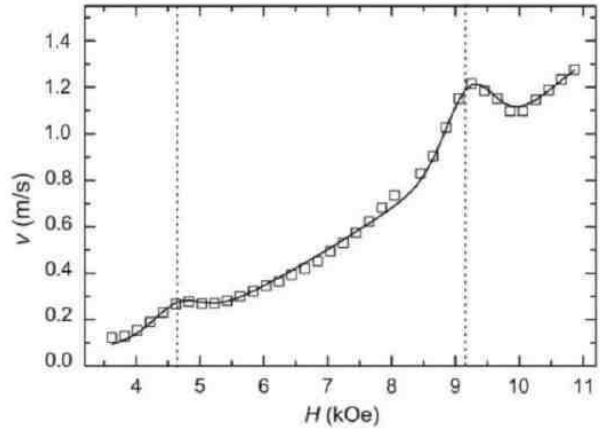}
\caption{The speed of propagation of the magnetic avalanche deflagration front is plotted as a function of the (fixed) field at which the avalanche is triggered; note the enhancement of propagation velocity at magnetic fields corresponding to quantum tunneling (denoted by vertical dotted lines).  From Hern\'andez-M\'inguez {\em et al.} \cite{quantumdeflagration}}
\label{Tejadamaxima}
\end{figure}

Following the initial studies of Suzuki {\em et al.} \cite{suzuki} in which avalanches were triggered stochastically in swept magnetic field, Hern\'andez-M\'inguez {\em et al.} \cite{quantumdeflagration,TejadaJMMM} carried out a systematic investigation of avalanche speeds as a function of a preset, constant magnetic field, $\mu_0 H_z$, for avalanches triggered controllably using surface acoustic waves.  From SQUID-based measurements of the total magnetization of a crystal of known dimensions, and the realization that the avalanche propagates as an interface between regions of opposite magnetization\cite{suzuki}, Hern\'andez-M\'inguez {\em et al.}  \cite{quantumdeflagration,TejadaJMMM} deduced the speed of propagation as a function of magnetic field shown in Fig. \ref{Tejadamaxima}.  The speed of the avalanches is enhanced at the resonant fields where tunneling occurs, confirming the important role of quantum mechanics, and prompting the authors to name the phenomenon ``quantum magnetic deflagration."  Similar results have been obtained from local, time-resolved magnetization measurements using micron-sized Hall sensors.

\begin{figure}[tb]
\centering
\includegraphics[width=4.5 in]{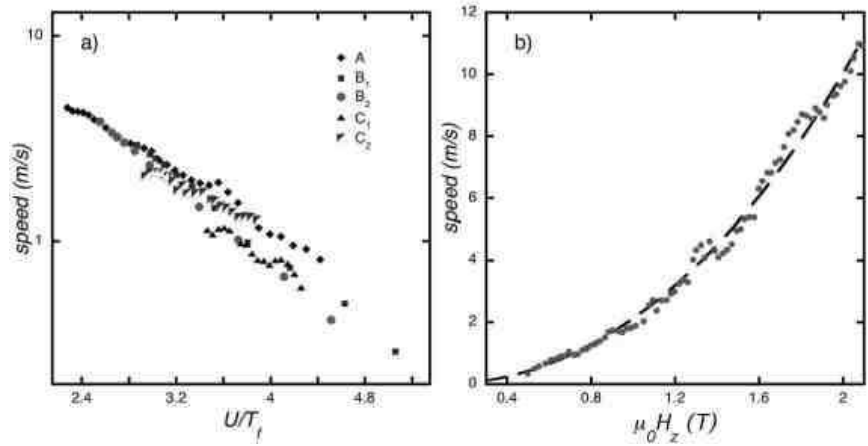}
\caption{(a) Avalanche speeds for a single crystal with various initial magnetic preparations.  $A$ denotes avalanches with $\Delta M/2M_s = 1$; $B_1$ and $B_2$ denote data taken at $\mu_0H_z = 2.2$ T  and $2.5$ T, respectively;  $C_1$ and $C_2$ denote avalanches with estimated flame temperatures $T_f \approx 10$ K  and $12$ K, respectively.  (b) Avalanche speeds for different crystal with $\Delta M/2M_s = 1$.  The fit requires an unphysical temperature dependence for the thermal diffusivity, $\kappa \propto T^{3.5}$.}
\label{tuning}
\end{figure}

McHugh et al. \cite{mchugh2} have reported a detailed, systematic investigation of the speed of magnetic avalanches for various experimental conditions.  The speed of propagation of an avalanche is described approximately \cite{suzuki} by the expression, 
$v\sim  (\kappa/\tau_0)^{1/2}$exp$[-U(H)/2k_BT_f]$, where $U$ is the barrier against spin reversal, $T_f$ is the flame temperature at or near the propagating front where energy is released by the reversing spins,  $\kappa$ is the thermal diffusivity, and $\tau_0$ is an attempt time.  In these studies, avalanches were controllably triggered: (A) in various external fields with fixed (maximum) initial magnetization, so that both $U$ and $T_f$ vary; (B) in fixed external field with different initial magnetization so that the avalanches differ primarily through the amount of energy released, and thus the flame temperature $T_f$; and (C) where external magnetic fields and initial magnetization are varied and adjusted to hold the energy released, and thus $T_f$, constant.  These parameters will be discussed in more detail in the next section.  We note at this point that the energy barrier $U$ and the flame temperature $T_f$ appear only as the ratio $U/T_f$ in the above expression for the velocity.  It is therefore convenient to plot the speed of the avalanche as a function of $U/T_f$ as is done in Fig. \ref{tuning}(a) \cite{mchugh2}.  

To end this section, we note that Villuendas et al. \cite{Villuendas} have recently reported a novel approach to studying avalanche dynamics using the magneto-optical Kerr effect to image the magnetization during an avalanche.  Although the resolution was quite limited, with further improvements this technique could provide a valuable avenue for studying magnetic deflagration.

\section{Comparison with the Theory of Magnetic Deflagration}
There are two essential ingredients for magnetic deflagration: the metastable spins release energy as they relax to the ground state, and this energy diffuses as heat through the crystal and thermally stimulates the reversal of neighboring metastable spins.  Garanin and Chudnovsky \cite{Garanin} developed a comprehensive theory of magnetic deflagration encapsulated in the following equations \cite{ferran}:
\begin{eqnarray}
C\frac{\partial T}{\partial t} -\nabla\cdot k \nabla T &=&  -\langle E\rangle \frac{\partial n}{\partial t}
\label{Diffusion}\\
\frac{\partial n}{\partial t} &=& -\Gamma n. 
\label{Rate}
\end{eqnarray}
Equation \ref{Diffusion} describes the flow of heat through the solid with the relaxing spins, $\frac{\partial n}{\partial t}$, as a source of heat; the thermal conductivity, $k$, is related to the heat capacity, $C$, through the thermal diffusivity, $\kappa$, as $k = \kappa C$.  Equation \ref{Rate} describes the relaxation of the metastable spin density, $n$, with a thermal relaxation rate $\Gamma$ given by an Arrhenius law, $\Gamma = \Gamma_0\mbox{ exp}[-U/k_BT]$; $\langle E\rangle$ is the average amount of heat released per molecule when its spin relaxes to the stable state.  For a single relaxing molecule, the energy change from the metastable to stable state is simply the Zeeman energy $\Delta E = 2g\mu_BSB_z$.  However, since not all spins necessarily relax during an avalanche, the average energy released per molecule must be introduced:
\begin{eqnarray}
\langle E \rangle = 2g\mu_BS \left( \frac{\Delta M}{2M_s}\right)B_z,
\label{DeltaE}
\end{eqnarray}
where $M_s$ is the saturation magnetization and $\Delta M = |M_z - M_s|$ is the change from initial to final magnetization.

Since both $k$ and $C$ are functions of temperature, Eqs. \ref{Diffusion} and \ref{Rate} are coupled, nonlinear differential equations, which make them very difficult to solve in general.  The samples used in the experiments generally have large aspect ratios, so that we can use simple approximations for quasi-one-dimensional avalanche propagation \cite{Garanin} to compare theory with experimental results.

We begin by comparing the theory with experimental results for the ignition temperatures and the stability criteria.  Heuristically, a deflagration front can develop when the rate at which energy is released by the relaxing metastable spins exceeds the rate of energy loss through the boundaries of the crystal.  This condition can be expressed in terms of a critical relaxation rate,\cite{Garanin} 
\begin{eqnarray}
\Gamma_c = \frac{8k(T_0)k_B T_0^2}{U\langle E\rangle l^2},
\label{criticalRate}
\end{eqnarray}
where $T_0$ is the initial temperature, $\Gamma_c = \Gamma_0\mbox{ exp}[-U/k_BT_0]$, and $l^2$ is the characteristic cross section of the crystal.  The curves shown in Fig. \ref{macia} (b) are the result of a calculation using Eq. \ref{criticalRate}. Two areas are defined in the $(H_z,H_x)$ parameter space where the spins are expected to be metastable against relaxation: the solid line denotes the region of metastability against relaxation by triggering avalanches while the dashed curve delineates the region of metastability against slow, stepwise relaxation.  If the experimental trajectory, denoted by the arrows, crosses the grey solid line first, an avalanche will ignite.  If the dashed line is crossed first, the metastable spins will relax slowly without triggering an avalanche.  This defines a critical angle $\theta_c$, above which an avalanche cannot occur.

Maci\`a {\em et al.} measured the ignition threshold by applying an increasing external field at an angle with respect to the crystal.  The relaxation rate increases as the field grows until $\Gamma_c$ is reached and deflagration ignites, as shown in Fig.~\ref{macia} (a).  For sufficiently large values of $H_x$, they found that the slow relaxation of the metastable spins occurs before deflagration can ignite.  This defines a line in parameter space separating regions where one or the other mode of relaxation occurs, as shown in Fig. \ref{macia}(b).  The theory predicts that the transverse field should result in a significant decrease in the magnetization metastability at the resonant fields of $H_z$.  The data recorded in Fig. \ref{macia}(a) confirm this and are consistent with the ignition temperatures of Fig. \ref{ignitionminima}.  In addition, ignition thresholds were measured at two different temperatures.  The area of stability is clearly reduced by the increased initial temperature, as expected.

To summarize, the theory of magnetic deflagration is in excellent agreement with the experiments of McHugh {\em et al.}, where the critical relaxation rate was reached by varying $T_0$ with a heater, and the experiments of Maci\`a {\em et al.}, where the ignition threshold was reached by controlling the barrier $U$ using $H_x$ and $H_z$.

We now proceed to compare the measured avalanche speeds with the theory of magnetic deflagration.  An approximate expression for the speed of the deflagration front is given by \cite{Garanin}
\begin{eqnarray}
v = \sqrt{\frac{3k_BT_f\kappa \Gamma(B, T_f)}{U(B)}}.
\label{OneDSpeed}
\end{eqnarray}
Eq. \ref{OneDSpeed} requires that the relaxation rate of the spins at the highest temperature, $\Gamma(U, T_f)$, is significantly slower than the rate at which heat traverses the interface width, $\kappa/\delta^2$, where $\delta$ is the magnetic interface width; this condition is expected to hold for all speeds considered here.

The barrier $U(B_z=\mu_0H_z)$ is calculated from the effective spin Hamiltonian (Eq. \ref{Hamiltonian}).  The temperature of the front, $T_f$, can be estimated from the measured heat capacity at zero magnetic field \cite{Gomes}, the calculated contribution of the spins in magnetic field, and the average energy released per molecule:
\begin{eqnarray}
\langle E\rangle = \int_{T_0}^{T_f}C(B_z, T)  dT. 
\label{Heat capacity}
\end{eqnarray}
Typical values calculated for $T_f$ range from $7$ K to $16$ K.  
 
With reference to the theoretical expression for the avalanche speed, Eq.~\ref{OneDSpeed}, if one assumes the thermal diffusivity $\kappa$ is approximately independent of temperature, or that its temperature dependence is unimportant compared to that of other parameters in the problem, then all measured avalanche velocities should collapse onto a single curve when plotted as a function of $[U(H)/T_f]$.  Although an approximate collapse is indeed obtained, as shown in  Fig. \ref{tuning} (a), there are clear and systematic deviations depending on whether (A): there is full (maximum) magnetization reversal, $\Delta M/2M_s = 1$ (both $U$ and $T_f$ vary); (B$_1$,B$_2$): the amount of ``fuel'' $\Delta M/2M$ is varied for a fixed magnetic field (thus $U$ is held constant); (C$_1$,C$_2$): avalanches are triggered such that the product $\Delta M \times H$ remains the same so that the energy released and the flame temperature $T_f$ are held constant.  That these different experimental protocols introduce systematic variations, albeit small, suggests that the theory is incomplete.

An attempt to fit to the theory by allowing the thermal diffusivity to vary with temperature as a power law is shown in Fig. \ref{tuning} (b) for avalanches of type (A) that involve full magnetization reversal.  Note that the enhancements of the velocity at certain values of magnetic field are associated with the tunneling resonances, in agreement with Fig.\ref{Tejadamaxima}.  The fit with Eq. \ref{OneDSpeed} is obtained for a thermal diffusivity that varies with temperature as $\kappa \sim T^{3.5}$.  This is distinctly unphysical, as the thermal diffusivity is generally a strongly decreasing function of temperature \cite{LowTemperaturePhysics} for these materials.  A similar analysis was performed \cite{minors} on avalanches of the minor species that yields an even steeper increase of $\kappa$ with temperature.  Regrettably, experimental measurement of the thermal diffusivity of Mn$_{12}$ are not available.
 
We conclude that, although the theory of deflagration agrees well with the measured conditions for the ignition of avalanches and provides a description of the avalanche velocity that is qualitatively correct, there are detailed discrepancies that suggest that additional factors need to be included in the theory to obtain good quantitative agreement with experiment.

\section{Outlook}

More experimental work is clearly needed.  Measurements of the thermal diffusivity would provide an important constraint on the theory, as would a reliable determination of the (local) temperature of the deflagration front.  Investigations of the influence of sample shape, size and quality would also be useful.  Spatial control of the avalanche ignition points, possibly by optical means, could provide important information.  Studies of the shape of the deflagration front, and its character (turbulent or laminar) would be particularly interesting. 

The possibility of observing detonation is intriguing.  Deflagration is but one type of combustion process.  Another, more violent type, is detonation, where heat spreads from the reaction front as a shock wave rather than by diffusion.  It is natural to ask whether crystals of molecular magnets can support the magnetic analog of chemical detonation.  Decelle et al. \cite{Decelle} have reported intriguing results hinting at this possibility using high external field sweep rates (4 kT/s).  The interpretation of these experiments is not entirely clear, and much work remains to be done.  

We close by noting that, to the degree that magnetic deflagration resembles chemical deflagration, the magnetic manifestation of this process offers some major advantages for the systematic study of chemical combustion.  The magnetic analog is non-destructive, reversible and continuously tunable using an external field.  Unlike the chemical process, it is a particularly interesting realization of deflagration in which quantum mechanical tunneling plays an important role.

\section*{Acknowledgements}

We thank Ferran Maci\`a and Jonathan R. Friedman for their careful reading of the manuscript and for constructive suggestions.  Support was provided by NSF Grant No. DMR-00451605.

\end{document}